\documentclass[twocolumn,prl,showpacs]{revtex4}

\usepackage{epsfig}
\usepackage{dcolumn}
\usepackage{bm}
\usepackage{amsmath}

\def\be{\begin{equation}}
\def\ee{\end{equation}}

\include{psfig}

\begin{document}
\author{F. Mallamace$^{1,2}$, S. H. Chen$^{1}$, A.Coniglio$^{3,6}$,
L. de Arcangelis$^{4,6}$, E. Del Gado$^{5,6}$ and A. Fierro$^{3,6}$}
\affiliation{$^{1}$ Department of Nuclear Engineering, Massachusetts Institute
of Technology, Cambridge MA 02139}
\affiliation{$^{2}$ Dipartimento di Fisica and INFM,
Universita' di Messina, I-98166, Messina, Italy.}
\affiliation{${}^3$ Dipartimento di Scienze Fisiche, Universit\`a di Napoli
"Federico II",\\ Complesso Universitario di Monte Sant'Angelo,
via Cintia 80126 Napoli, Italy}
\affiliation{${}^4$Dipartimento di Ingegneria dell'Informazione,
 Seconda Universit\`a di Napoli, via Roma 29, 81031 Aversa (Caserta), Italy}
\affiliation{${}^5$ Laboratoire des Verres, Universit\'e Montpellier II, 34095
Montpellier, France}
\affiliation{${}^6$ INFM Coherentia and Udr di Napoli, Napoli, Italy}

\title{{\bf Complex viscosity behavior and cluster formation in attractive
colloidal systems }}

\begin{abstract}
The increase of the viscosity, which is observed in attractive colloidal
systems by varying the temperature or the volume fraction, can be related
to the formation of structures due to particle aggregation.
In particular we have studied the non trivial dependence of the viscosity from
the temperature and the volume fraction in the copolymer-micellar system $L64$.
The comparison of the experimental data with the results of numerical
simulations in a simple model for gelation phenomena suggests that this
intriguing behavior can be explained in terms of cluster formation and that
this picture can be quite generally extended to other attractive colloidal
systems.

\end{abstract}

\date{\today}

\maketitle

\smallskip%
%

Soft materials (polymers, proteins, colloids, etc.) represent a research field
of large interest in science and technology and are a common domain for
physics, chemistry and biology.
The effective interaction amongst particles in these systems is in many cases
the result of the competition between repulsive and attractive contributions
and may give rise to aggregation processes with the formation of a variety of
meta-stable or stable mesoscopic structures \cite{nature_clu}-\cite{science}.
This is the case of gelation phenomena in attractive colloidal systems at
low volume fractions \cite{weitz1}-\cite{weitzn}. 
Here slow dynamics and eventually structural arrest are
associated to the formation of long-living non-compact structures, with a
change in the viscoelastic properties as the one typically observed in 
gelation phenomena.
At present, this is a point of central interest in the study of
attractive colloidal systems as the effects of aggregation processes at low
volume fractions on the dynamics are still poorly understood \cite{cates}-
\cite{jpc}.

A first interesting clue comes from the better known gel formation in polymer
systems: here the phenomenology of gelation is typically interpreted in
terms of a percolation transition \cite{flory}-\cite{coniglio}, 
that is the formation of a permanent spanning structure, accompanied by a 
structural arrest.
This suggests that gelation phenomena
could be understood in terms of formation of clusters, and
eventually of a spanning structure, also in attractive colloidal systems.
Here, on the other hand, at high volume fractions and low temperatures,
structural arrest phenomena are observed with typical glassy
features, that have been interpreted within the framework of the mode coupling
theory \cite{mct1,mct2,mct3}.
In this case the glassy behavior can be strongly influenced by the short range
attraction, but it cannot be simply described in terms of
formation of clusters or of a percolating structure.
Therefore the role of structure formation in the slow
dynamics and the structural arrest of attractive colloidal systems is not
completely clear.

In this work we discuss the viscosity behavior in the copolymer
micellar system $L64$, where the short range attraction is due to
an effective intermicellar interaction. We observe a
transient gelation phenomenon followed by a structural arrest.
For a fixed value of the temperature, as function of the volume fraction the 
viscosity increases, reaches a plateau and then increases again at 
higher volume fraction. In this process the rheological properties exhibit a 
crossover from gelation to a structural arrest typical of the glass 
transition.
The comparison of the experimental findings
with the numerical study of a simple model suggests that these two
phenomena, which appear to be quite distinct, stem from two
different mechanism: One related to the presence of long living clusters, 
the other due to crowding of particles. 
The model also suggests that the interrupted
gel phenomena can be more generally related to colloidal
gelation \cite{weitz1}-\cite{bartlett}. 
In other colloidal systems, due to different features of the effective
attractive interaction, the gelation phenomenon occurs at much lower 
temperature and volume fraction and it is more pronounced. 
Consequently, structural arrest and gelation
almost coincide and occur close to the percolation transition. 
At higher temperatures clusters can be less persistent and
structural arrest, occurring at higher volume fractions, exhibits features of 
the so called attractive glasses. 
Increasing further the temperature the effect of the bonding interaction 
vanishes and one finds features of the hard sphere glass transition
\cite{mct1,mct2,mct3}.

\begin{figure}[ht]
\begin{center}
\mbox{\epsfxsize=8cm\epsfysize=8cm\epsfbox{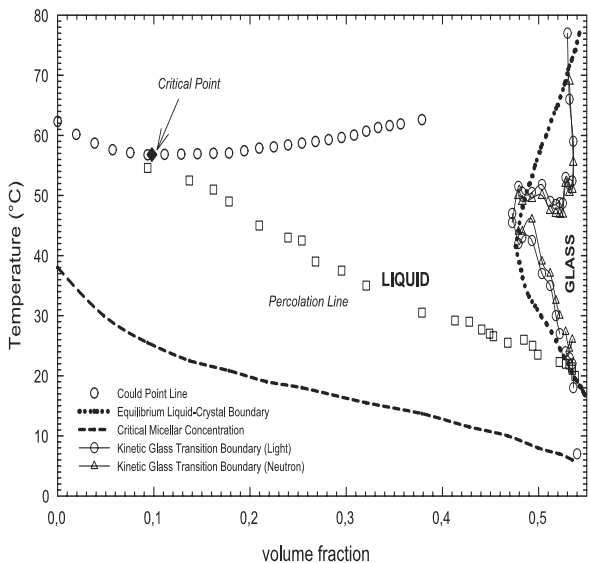}}
\end{center}
\caption{The phase diagram of the $L64/D_{2}O$ system (adapted from 
Ref.~\cite{sans1}): the inverted binodal line (circles), the percolation line 
(empty squares) and the glassy line (circles and triangles on the right of the
figure). 
The percolation and the glassy lines were measured by both
quasi elastic light scattering and small angle neutron scattering.}
\label{phasediagram}
\end{figure}

We have studied the viscosity behavior, as a function of the temperature
$T$ and of the micellar volume fraction $\Phi$, in the copolymer-micellar
system $L64$ \cite{lobry}-\cite{nos1}. 
This is an aqueous solution of a non-ionic
three-block copolymer made of polyethylene and polypropylene
oxides, $[(PEO)_{13}-(PPO)_{30}-(PEO)_{13}]$ ($L64$ for short).
These polymer molecules in water are $T$ dependent surfactants
that spontaneously form hydrated monodisperse spherical micelles
in a wide $T-\Phi$ range. At low temperatures, both $PEO$ and
$PPO$ are hydrophilic, so that the $L64$ chains readily dissolve
in water, and the polymer molecules exist as unimers. However,
increasing $T$, $PPO$ tends to become less hydrophilic than $PEO$
creating an unbalance in the hydrophilicity between the central
block and the end block of the polymer molecule that consequently
acquires surfactant properties and self-assemble to form micelles.
The inter-particle attraction is mainly due to the fact that, on
further increasing $T$, water becomes progressively a poor solvent
to both $PEO$ and $PPO$ chains. The phase diagram of the $L64/D_{2}O$ system
is shown in Fig.~\ref{phasediagram}: Due to
the intermicellar attraction, the system has an inverted binodal curve with a
lower consolute critical point, a $T-\Phi$ dependent percolation
line where the viscosity exhibits a steep increase,
and a glassy line as predicted by the mode coupling theory
\cite{lobry,sans1}. In particular, by increasing the volume fraction the
percolation line precedes the glassy line.

Viscosity measurements, as a function of $T$ at different $\Phi$, were
performed with a strain-controlled rheometer using a double wall Couette
geometry, at the fixed frequency $\omega =1$ $\sec ^{-1}$. To ensure a
linear response, for different $T$  and $\Phi$, we have tested
our system as a function of low applied strains, $\gamma$. After this check
a low strain deformation, $\gamma =0.05$, was maintained for all the
experiments.  A crucial point is that the Peclet number
$Pe=\overset{.}{\gamma }\xi ^{2}/2D_{0}$, which characterizes the
amount of distortion of structures with linear dimension $\xi $ ($D_{0}$ is
the particle short-time diffusion coefficient at low concentration where
hydrodynamic effects dominate) must be small.
In Ref.~\cite{peclet} it was found that $\xi \sim 5000
\overset{o}A$ and $D_{0}$ of the order of $2 \cdot 10 e^{-11} m^{2}/s$ 
close to the percolation threshold, ensuring $Pe < 1$.
These distortions are 
responsible for shear
thinning in the system structures, which is expected to set on at $Pe\approx
1$. In other words, the condition $Pe<1$ constitutes an important experimental
constraint in the study of the elasticity of systems characterized by
clustering processes like polymer solutions and attractive colloids.
In particular, this is true for viscosity measurements in phenomena dominated
by long-range ``critical'' correlations, like
those which originate the diverging incipient cluster typical of the
percolation threshold in the sol-gel transition. Just in these processes,
shear thinning can originate a ``levelling
off'' of the measured viscosities. 

\begin{figure}[ht]
\begin{center}
\mbox{\epsfxsize=8cm\epsfysize=8cm\epsfbox{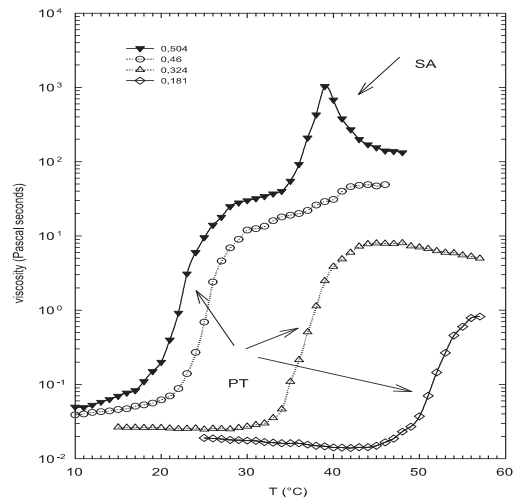}}
\end{center}
\caption{ The $L64/D_{2}O$ viscosity, $\eta$, as a function of
the temperature, $T$, for different values of the volume fraction $\Phi$.
In figure the percolation transition (PT) and the structural arrest (SA) are
shown.}
\label{figura1}
\end{figure}

In Fig.~\ref{figura1} the measured viscosity, $\eta$, of the
system is plotted as a function of the temperature, $T$,
at different volume fractions, $\Phi$. 
At each $\Phi$, $\eta$ is characterized by a
growth from the values typical of the solvent ($\eta \sim
10^{-2}Pa$) followed by a bending at high temperatures.
At high $\Phi$ a second growth is observed in $\eta$ by further increasing the
temperature:
The first growth of the viscosity corresponds to the percolation transition (PT)
\cite{lobry}, and the second growth to
the structural arrest (SA) of the attractive glass transition \cite{nosjpc}.

\begin{figure}[ht]
\begin{center}
\mbox{\epsfxsize=8cm\epsfysize=8cm\epsfbox{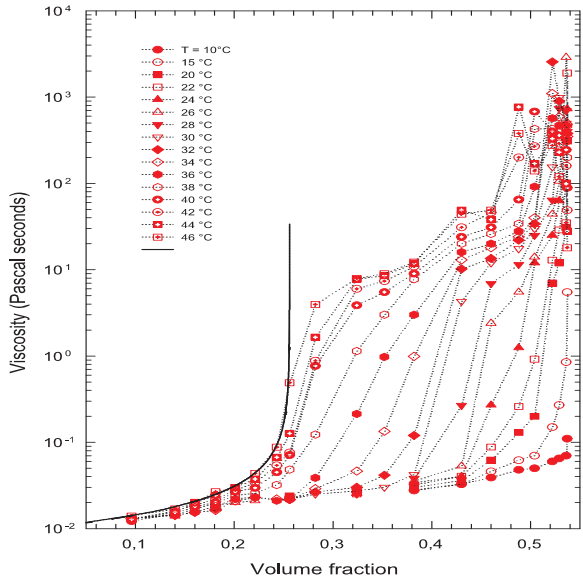}}
\end{center}
\caption{ The $L64/D_{2}O$ viscosity, $\eta$, as a
function of the volume fraction, $\Phi$, for different values of the
temperatures, $T$, in the range $10-46
{{}^\circ}%
C.$ The continuous line is a power law, $(\Phi_c-\Phi)^{-k}$, with $k=1.35$.
}
\label{figura2}
\end{figure}

In Fig.~\ref{figura2} $\eta$ is plotted as a
function of the volume fraction, $\Phi$, for different values of the
temperature, $T$.
At low volume fraction a first strong growth of the viscosity is found close to
the percolation line.
However, a bending in the viscosity is found, as the volume fraction is
increased above the percolation threshold,
and finally,
by further increasing the volume fraction, a second growth corresponding to the
glass transition \cite{nosjpc} is observed.
A similar effect in the viscosity behavior has also been observed in
another micellar system in Ref.~\cite{durand}.

The strong growth of the viscosity close to the percolation line
is reminiscent of the viscosity divergence observed in polymer
gelation, and might be due to the presence of long living clusters
in the system. The intermicellar interaction has an entropic origin. In fact, 
the micelle can be schematized as a hard sphere with polymers attached on the 
surface. When two micelles come close, the polymers get entangled 
(long-living bonds). Due to the entanglement,
the two micelles have to overcome an entropic barrier to get separated.
Therefore the bond lifetime $\tau_b$ will be related to
the height of such entropic barrier $S_b$, i.e. $\tau_b \sim e^{S_b}$ (the
entropy is in units of Boltzmann's constant). The
other quantity of interest is the characteristic time $\tau_u $ for
two close micelles to get entangled (bonded): $\tau_u\sim e^{S_u}$
where $S_u$ is the entropic barrier needed to be overcome to go
from the un-bonded state to a bonded state.
Therefore the probability $p_b$
for two close micelles to be bonded is given by $$p_b =\frac{e^{S_b}}
{e^{S_b} + e^{S_u}}.$$ 
Note that such bond
probability can be quite smaller than $1$, depending on the value
of $S_u$. This could explain 
why in micellar systems percolation can occur at quite high volume fraction, 
as compared to other colloidal systems \cite{weitz1}-\cite{weitzn}.

To study the effect of aggregation and structural arrest in such
micellar system, we propose a model which contains these two
characteristic times for forming and breaking bonds. Tuning
the parameters $\tau_b$ and $\tau_u$ we can study the gelation process
with infinite bond lifetime (polymer gelation) and finite bond
lifetime (micelles). A detailed discussion on the model and on the
numerical simulations can be found in Ref.~\cite{concon}.

We consider a solution of tetrafunctional monomers with excluded
volume interactions and performed Monte Carlo simulations on the cubic lattice.
At $t=0$ the volume fraction $\Phi$ is fixed,
and bonds between monomers are randomly quenched. The four possible
bonds per monomers, randomly selected, are formed 
along lattice directions between monomers that are nearest neighbors and
next nearest neighbors.
In the case of polymer gelation, once formed the
bonds are permanent. The monomer diffuse according to the bond-fluctuation
dynamics (BFD)\cite{18con}. We let the monomer diffuse to reach the stationary
state and then study the system for different values of the concentration.
The system has a percolation transition
at $\Phi _{c}=0.376\pm 0.003$. 
In the case of colloidal gelation, we introduce a finite bond lifetime
$\tau_{b}$. We start with the same
configurations of the previous case, with a fixed $\Phi$ where the bonds have
been randomly quenched. During the monomer diffusion with
$BFD$ at each time step we attempt to break each bond with a frequency
$1/\tau_{b}$. Between monomers separated by a distance less than $l_{0}$ bonds
are then formed with a frequency $1/\tau_{u}$.

From the  time autocorrelation function, $f_{\vec{q}}(t)$, of equilibrium 
density fluctuations with 
$\vec{q}\equiv (\frac{\pi}{4},\frac{\pi}{4},\frac{\pi}{4})$, we calculate 
the relaxation time $\tau$ as the time
such that $f_{\vec{q}}(\tau) \sim 0.1$. 
\begin{figure}[ht]
\begin{center}
\mbox{\epsfxsize=8cm\epsfysize=8cm\epsfbox{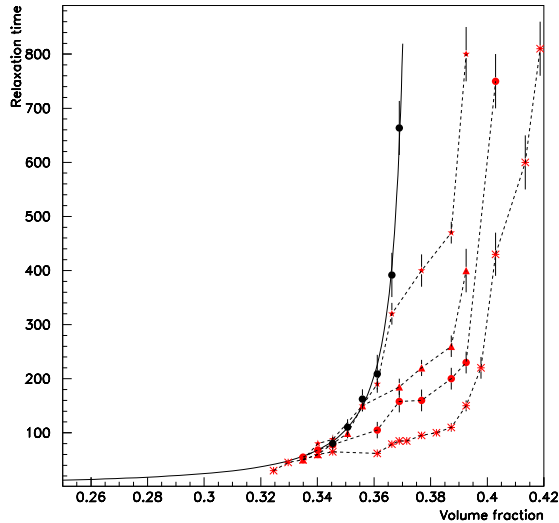}}
\end{center}
\caption{The average relaxation time, $\tau$, as a function of the volume
fraction, $\Phi$.
From left to right: the data for the permanent bond case
diverge at the percolation threshold with a power law (continuous line),
see text. The other data refer to finite 
$\tau _{b}=3000,1000,400,100$ $MCstep/particle$
decreasing from left to right (the dashed lines are guides to the eye).
(Adapted from Ref.~\cite{concon}).}
\label{figura4}
\end{figure}
In Fig.~\ref{figura4}
the relaxation time, $\tau$, is plotted as function of the
volume fraction, $\Phi$, for the permanent bonds and for the
finite lifetime bonds case at different values of $\tau_b$. In the figure
one finds the data for the permanent bond case on the left, and from left
to right the data for finite bond lifetime, for
decreasing values of $\tau_{b}$.

In the case of permanent bonds, $\tau(\Phi)$
displays a power law divergence, $(\Phi_c-\Phi)^{-k}$, at the
percolation threshold, with $k=1.35$. 
For finite bond lifetime $\tau_{b}$ the relaxation time instead
increases following the permanent bond case, up to some value
$\Phi^{*}$ and then deviates from it. The longer the bond lifetime
the higher $\Phi^{*}$ is. For higher $\Phi$ the increase of the
relaxation time corresponds to the onset of the glassy regime in
the relaxation behavior discussed in Ref.~\cite{concon}. 
This qualitative behavior has also been confirmed via molecular dynamics 
simulations in Ref.~\cite{nota_sciort}.  
This truncated critical behavior of the
relaxation time qualitatively reproduces the experimental data shown in
Fig.~\ref{figura2}. In order to stress the analogy, we have plotted a power
law behavior for the viscosity with the same value of $k=1.35$ (continuous 
curve in Fig.~\ref{figura2}).

This comparison suggests the following interpretation for the experimental data.
We consider that clusters of different sizes will be present in the
system at any time $t$.
In the permanent bond case, a cluster of radius $R$ diffuses amongst the others
with a characteristic relaxation time, $\tau(R)$: At the percolation
threshold the connectedness length critically grows in the system and so does
the overall relaxation time.
In the case of a finite bond lifetime, $\tau_{b}$, there will be a cluster size
$R^{*}$ so that $\tau_{b} \sim \tau(R^{*})$. That is, clusters of size
$R \leq R^{*}$ can be considered as permanent, whereas clusters of size 
$R >  R^{*}$ will not contribute with their full size to the enhancement
of the relaxation time in the system.
Therefore, a long but finite bond lifetime produces the enhancement of the
viscosity due to the formation of persistent clusters. On the other hand,
it introduces an effective cluster size distribution with a cut-off which
keeps the macroscopic viscosity finite in the system \cite{conmess}.

In conclusions, based on a minimal model for colloidal system
we have analyzed  the experimental viscosity data
in copolymer-micellar systems. For a fixed
temperature the viscosity exhibits a first steep growth as the
volume fraction increases. This effect is due to coalescing clusters of
monomers linked by long living bonds. Thus the viscosity increases as
in a typical gelation phenomenon, but due to the finite
lifetime of the bonds this increase is interrupted and
followed by a saturation. As the volume fraction increases
further, the viscosity exhibits a second steep growth until structural arrest
is reached. The resulting glass transition can be related to the formation of
a spanning cluster made of localized particles connected by bonds.
It is the presence of the bonds which makes the difference between the
attractive glass and a repulsive glass.

The research in Italy is supported by EU Network Number
MRTN-CT-2003-504712, MIUR-PRIN 2004, MIUR-FIRB 2001, CRdC-AMRA,
INFM-PCI. The research at MIT is supported by grant from Material
Science Division of USA-DOE. We also acknowledge the EU-Marie
Curie Individual Fellowship HPMF-CI2002-01945.

%

\end{document}